\documentclass{elsart3p}
\usepackage{times}
\usepackage[numbers,sort&compress]{natbib}
\usepackage{amssymb,amsmath}
\usepackage[final]{graphicx}
\usepackage{units}
\usepackage{color,psfrag}

\newcommand{\Lc}{\mathcal L}
\newcommand{\Mpl}{m_\mathrm{Pl}}
\newcommand{\mPl}{m_\mathrm{Pl}}
\begin{document}

%%%%%%%%%%%%%%%%%%%%%%%%%%%%%%%%%%%%%%%%%%%%%%%%%%%%%%%%%%%%%%%%%%%%%%%%
\begin{frontmatter}
\title{Inflaton mass in the $\nu$MSM inflation}

\author[DESY]{Alexey Anisimov},
\ead{alexey.anisimov@desy.de}
\author[EPFL]{Yannick Bartocci},
\author[MPI,EPFL,INR]{Fedor L. Bezrukov}
\ead{Fedor.Bezrukov@mpi-hd.mpg.de}

\address[DESY]{
  Notkestrasse 85,
  DESY T, Build 2a
  22607 Hamburg,
  Germany}
\address[EPFL]{
  Institut de Th\'eorie des Ph\'enom\`enes Physiques,
  \'Ecole Polytechnique F\'ed\'erale de Lausanne,
  CH-1015 Lausanne, Switzerland}
\address[MPI]{
  Max-Planck-Institut f\"ur Kernphysik,
  PO Box 103980,
  69029 Heidelberg,
  Germany}
\address[INR]{
  Institute for Nuclear Research of Russian Academy of Sciences,
  Prospect 60-letiya Oktyabrya 7a,
  Moscow 117312, Russia}

\date{19 December 2008}

\begin{abstract}
  We analyze the reheating in the modification of the $\nu$MSM
  (Standard Model with three right handed neutrinos with masses below
  the electroweak scale) where one of the sterile neutrinos, which
  provides the Dark Matter, is generated in decays of the additional
  inflaton field.  We deduce that due to rather inefficient transfer
  of energy from the inflaton to the Standard Model sector reheating
  tends to occur at very low temperature, thus providing strict bounds
  on the coupling between the inflaton and the Higgs particles.  This
  in turn translates to the bound on the inflaton mass, which appears
  to be very light $\unit[0.1]{GeV}\lesssim m_I\lesssim
  \unit[10]{GeV}$, or slightly heavier then two Higgs masses
  $\unit[300]{GeV}\lesssim m_I\lesssim \unit[1000]{GeV}$.
\end{abstract}

\begin{keyword}
  Inflation \sep Standard Model \sep reheating \sep nuMSM
  %\PACS 98.80.Cq \sep 14.80.Bn
  % 98.80.Cq Particle-theory and field-theory models of the early
  % Universe (including cosmic pancakes, cosmic strings, chaotic
  % phenomena, inflationary universe, etc.)
  % 14.80.Bn 	Standard-model Higgs bosons
\end{keyword}

\end{frontmatter}
%%%%%%%%%%%%%%%%%%%%%%%%%%%%%%%%%%%%%%%%%%%%%%%%%%%%%%%%%%%%%%%%%%%%%%%%

%%%%%%%%%%%%%%%%%%%%%%%%%%%%%%%%%%%%%%%%%%%%%%%%%%%%%%%%%%%%%%%%%%%%%%%%
\section{Introduction}

In \cite{Asaka:2005an,Asaka:2005pn} it was shown that within the
Standard Model (SM) complimented with three right-handed neutrinos
$N_I$ with the masses smaller than the electroweak scale one can
simultaneously explain both the dark matter and the baryon asymmetry
of the universe
\cite{Asaka:2005an,Asaka:2005pn,Bezrukov:2005mx,Boyarsky:2006jm,%
  Asaka:2006ek,Shaposhnikov:2006xi,Shaposhnikov:2006nn,Asaka:2006rw,%
  Asaka:2006nq,Bezrukov:2006cy,Gorbunov:2007ak,Shaposhnikov:2007nj}.
This model dubbed as $\nu$MSM represents a particular realization of
the seesaw extension of the SM and is fully consistent with the
current experimental data from the light neutrino sector.  However,
generation of the proper Dark Matter abundance of the sterile neutrino
is not simple during the thermal evolution of the Universe, and
requires some amount of fine-tuning
\cite{Shaposhnikov:2008pf,Laine:2008pg}.  Being very weakly coupled,
sterile neutrinos do not reach thermal equilibrium, so an interesting
possibility is to generate them before the beginning of the standard
thermal history.  In \cite{Shaposhnikov:2006xi} such mechanism was
proposed, where the $\nu$MSM model was extended by adding the inflaton
field, which generates all the masses in the model and decays into
the SM particles \emph{and} sterile neutrinos after inflation,
\begin{align}
  \Lc_{\nu{\rm MSM}}\to \, &(\Lc_{\nu{\rm MSM}[M_I\to 0]}-
  {f_I\over 2}{\bar N}^c_IN_I\chi+{\rm h.c.})+
  \notag\\
  &{1\over
    2}(\partial_{\mu}\chi)^2+|\partial_{\mu}\Phi|^2-V(\Phi,\chi)
  \;,
  \label{lagr}
\end{align}
where $\Phi$ and $\chi$ are the Higgs and the inflaton fields
correspondingly and
\begin{multline}
  \Lc_{\nu{\rm MSM}[M_I\to 0]}=
  \Lc_{MSM}+{\bar N}_Ii\partial_{\mu}\gamma^{\mu}N_I\\
  -F_{\alpha I}{\bar L}_{\alpha}N_I\Phi+\mathrm{h.c.}
\end{multline}
is the $\nu$MSM Lagrangian with all the dimensional parameters being
put to zero.  The potential $V(\Phi,\chi)$ is\footnote{In order to
  avoid the domain wall problem a cubic term $\mu\chi^3$ can be
  introduced.  It will be further assumed that $\mu\lesssim
  \sqrt{\alpha^3/\lambda}\,v_\mathrm{EW}$.  In that case such term has
  no influence on the dynamics of the model during the reheating
  stage, and the relation (\ref{mhi}) for the values of the parameters
  considered in the Letter is not altered significantly either.}
\begin{multline}
  \label{pot}
  V(\Phi,\chi) =
  \lambda\left(\Phi^{\dagger}\Phi-{\alpha\over\lambda}\chi^2\right)^2
  +{\beta\over4}\chi^4\\
  -{1\over 2}m^2_{\chi}\chi^2+V_0
  \;,
\end{multline}
where $V_0={m^4_{\chi}\over 4\beta}$ was introduced in order to cancel
the vacuum energy.  Expanding (\ref{pot}) around its vacuum
expectation value one has the relation between the
inflaton\footnote{Notations $I$ and $H$ will be used throughout the
  Letter to represent the diagonalized excitations above the vacuum
  expectation value for \eqref{pot}.  $I$ is the one mostly mixed with
  inflaton $\chi$, and $H$ mostly mixed with the SM Higgs $\Phi$.}
mass $m_I$ and the Higgs mass $m_H$:
\begin{equation}
  \label{mhi}
  m_I=m_H\sqrt{\beta\over 2\alpha}
  \;.
\end{equation}
If $\alpha>\beta/2$ the inflaton mass is smaller then the Higgs mass
and, therefore, the decay of the inflaton into the Higgs can only
occur in a thermal bath. In what follows we will first concentrate on
this particular case.  Parameter $\beta$ is fixed by the COBE
normalization of the amplitude of scalar perturbations
\cite{Lyth:1998xn}, $\beta\simeq 1.3\times 10^{-13}$.  Pure quartic
potential inflation is currently disfavored by the WMAP5 data
\cite{Spergel:2006hy} because of the too large predicted value of the
tensor to scalar amplitudes ratio.  However, if one allows non-minimal
coupling of the inflaton to gravity \cite{Tsujikawa:2004my} one can
bring this potential in agreement with the data.  This, in turn, will
influence the bounds on the inflaton mass.  We will discuss this in
the end of the Letter.

The upper constraint on the value of $\alpha$ comes from the
requirement that radiative corrections do not spoil the flatness of
the inflaton potential and is given by $\alpha\le 3\times 10^{-7}$.
This corresponds to the lower bound on the inflaton mass
\begin{equation}
  m_I\ge 0.07\left(m_{H}\over \unit[150]{GeV}\right)\sqrt{\beta\over
    1.3\times 10^{-13}}\,\mathrm{GeV}
  \;.
\end{equation}
One should note that larger values of $\alpha$ (leading to smaller
inflaton masses) may also be possible, but then the analysis of the
loop corrections to the effective potential of the inflaton becomes
important.

The lower bound on $\alpha$ comes from the requirement to have
successful baryogenesis in $\nu$MSM \cite{Asaka:2005pn}.  To allow for
efficient sphaleron conversion of the lepton asymmetry to baryon
asymmetry requires the reheating temperature to be larger then roughly
$\unit[150]{GeV}$ \cite{Burnier:2005hp}.  In
\cite{Shaposhnikov:2006xi} it was advocated that the resulting lower
bound is $\alpha>\beta\sim 10^{-13}$.  Below we will argue that the
lower bound is quite a bit stronger which leads to a narrow window for
the inflaton mass.

%%%%%%%%%%%%%%%%%%%%%%%%%%%%%%%%%%%%%%%%%%%%%%%%%%%%%%%%%%%%%%%%%%%%%%%%
\section{Reheating bounds}

Reheating after inflation proceeds through a regime of the parametric
resonance.  The dynamics of the models with potentials similar to
(\ref{pot}) in the parametric resonance regime was studied \emph{via}
analytic methods in, e.g.\ \cite{Greene:1997fu,Kofman:1997yn}.  The
analysis of the late stages of \emph{preheating} was made possible
with the lattice simulations package LatticeEasy
\cite{Micha2003,Micha2004,Felder:2000hq,Felder:2007nz}.  In
particular, the \emph{preheating} in the model with the potential
which contains only first two terms in (\ref{pot}) have been studied
in \cite{Micha2004}.

\begin{figure}
  \psfrag{1000}[bl]{\scriptsize{$10^{3}$}}
  \psfrag{t}[l]{$t_{pr}$}
  \psfrag{c}[bl]{$n_{\chi}$}
  \psfrag{p}[bl]{$n_{\phi}$}
  \psfrag{h}[bl]{\footnotesize{  $\alpha$ =$10^{-9}$}}
  \psfrag{i}[bl]{\footnotesize{  $\alpha$ =$10^{-10}$}}
  \psfrag{j}[bl]{\footnotesize{  $\alpha$ =$10^{-11}$}}
  \psfrag{k}[bl]{\footnotesize{  $\alpha$ =$10^{-12}$}}
  \psfrag{l}[bl]{\footnotesize{  $\alpha$ =$10^{-13}$}}
  \includegraphics[width=0.88\linewidth]{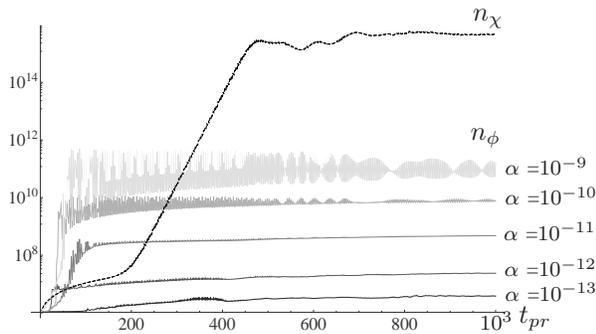}
  \caption{Number densities of the Higgs boson and the inflaton are
    shown for different values inflaton-Higgs coupling $\alpha$. Higgs
    self-coupling is taken as $\lambda=10^{-2}$. Time is given in
    program units, see \cite{Felder:2000hq}.  Preheating ends earlier
    for Higgs field ($t_{pr}\lesssim 100$) than for inflaton ($t_{pr}\lesssim
    500$).  For $\alpha=10^{-9}$ one has the border case when the
    average momenta of the fields are less then the lattice
    ultraviolet cutoff.}
  \label{fig1}
\end{figure}

At large values of the inflaton field $\chi$ the behavior is that of
the pure quartic inflation.  The expectation value of the Higgs field
$\Phi$ is set along the flat direction:
$|\Phi|^2={\alpha\over\lambda}\chi^2$.  After the end of inflationary
slow roll regime the inflaton field starts to oscillate.  In the very
beginning all the energy is stored in the zero (or homogeneous) mode
of the inflaton $\chi_0$, and all other modes are absent.  The
oscillations of $\chi_0$ initially excites the nonzero modes of both
the Higgs and the inflaton.  One can compare the contribution of the
zero mode of the inflaton to the effective masses of the Higgs and the
inflaton:
\begin{equation}
  m^2_{\mathrm{eff},\Phi}\sim \alpha\chi^2_0
  \;,\quad
  m^2_{\mathrm{eff},\chi}\sim
  \beta\chi^2_0
  \;.
\end{equation}
If $\alpha>\beta$ the corresponding contribution to the effective mass
of the Higgs is larger. Therefore at early stages of the evolution the
energy transfer into the Higgs particles is the dominating process.
This is in accord with \cite{Greene:1997fu,Kofman:1997yn}, and can be
inferred from the early time behavior of the number densities shown in
Fig.~\ref{fig1}.  One could then expect that the whole energy of the
inflaton field will be transferred exponentially fast to the Higgs
particles.\footnote{One can easily verify that only a small fraction
  of the energy of the inflaton is drawn into sterile neutrinos
  because of the smallness of the Yukawa couplings $f_I$ (at most
  $\sim10^{-7}$ for the heaviest sterile neutrinos lighter, then the
  inflaton). In particular, the rate $\Gamma(I\to NN)$ typically
  equilibrates at the temperatures below the temperature of the
  electroweak phase transition. The process involving the SM Yukawa
  couplings $F_{\alpha I}$ proceeds via the Higgs particle, and is
  even more suppressed.}  Since the Higgs decay to the SM fields and
their consequent thermalization are fairly fast compared to the Hubble
rate one could then estimate the resulting reheating temperature as in
\cite{Shaposhnikov:2006xi}
\begin{equation}
  T_R\sim m_{\rm Pl}\left(\alpha^2\over g_*\lambda\right)^{1\over 4}
  \;,
\end{equation}
which for $\lambda\sim 0.1$, the number of the SM d.o.f.\ $g_*\sim
10^2$ and $\alpha>\beta$ leads to the values of $T_R$ which greatly
exceed the freeze-out temperature of the sphaleron processes.

%%%%%%%%%%%%%%%%%%%%%%%%%%%%%%%%%%%%%%%%%%%%%%%%%%%%%%%%%%%%%%%%%%%%%%%%
%%%%%%%%%%%%%%%%%%%%%%%%%%%%%%%%%%%%%%%%%%%%%%%%%%%%%%%%%%%%%%%%%%%%%%%%

\begin{figure}
  \begin{center}
    \psfrag{l}[bl]{$\lambda$}
    \psfrag{vert}[bl]{$\rho_{\phi}/\rho_{\chi}$}
    \psfrag{txt}[l]{$\frac{\rho_{\phi}}{\rho_{\chi}}\propto \lambda^{-1}$}
    \psfrag{0.1}[bc]{\scriptsize{    $10^{-1}$}}
    \psfrag{0.001}[bl]{\scriptsize{$10^{-3}$}}
    \includegraphics[width=1.\linewidth]{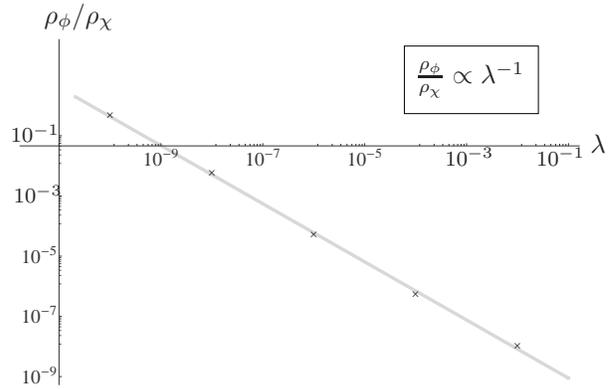}
  \end{center}
  \caption{Energy transfer dependence on $\lambda$ (here
    $\alpha=\beta=2.6\times10^{-13}$). Values are taken at late time
    $t_{pr}=10^3$. LatticeEasy parameters are as in Fig.~\ref{fig1}.}
  \label{fig2}
\end{figure}

The lattice results, however, show that such exponential energy
transfer into the Higgs particles for a broad range of parameters
terminates before any significant part of the inflaton zero mode
energy is depleted.  The reason for that is the large Higgs boson
self-coupling $\lambda\sim 0.1$ which makes the re-scattering
processes become important quite early. Unless the Higgs-inflaton
coupling $\alpha$ is fairly large the re-scatterings terminate the
resonance when only a negligible part of the energy in the inflaton
zero mode is depleted.\footnote{For a potential without the inflaton
  mass term in a different part of the parameter space similar claims
  were made in \cite{Micha2004}.}  On Fig.~\ref{fig2} one can see how
the amount of the transferred energy depends on the value of the Higgs
self coupling $\lambda$ which we allowed to vary to small values just
to demonstrate the importance of the re-scattering processes.

\begin{figure}
  \begin{center}
    \psfrag{a}[bl]{$\alpha$}
    \psfrag{t}[l]{$\frac{\rho_{\phi}}{\rho_{\chi}}\propto\alpha^{1.6}$}
    \psfrag{0.01}[bc]{\textcolor[rgb]{1.00,1.00,1.00}{l
                                        l     }\scriptsize{$10^{-2}$}}
    \psfrag{x}[l]{$\frac{n_{\phi}}{n_{\chi}}\propto \alpha^{1.3}$}
    \psfrag{b}[bl]{\footnotesize{$3\times10^{-8}$}}
    \includegraphics[width=1.\linewidth]{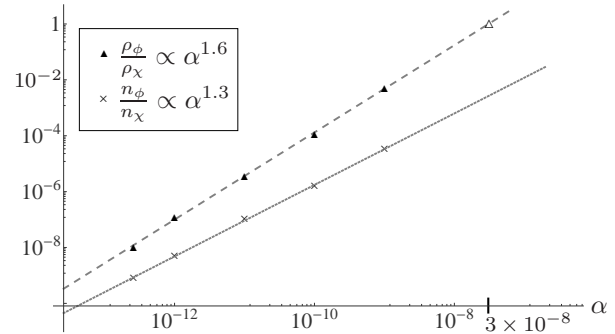}
  \end{center}
  \caption{Energy transfer dependence on inflaton--Higgs coupling
    $\alpha$. Values are taken at late time $t_{pr}=10^3$. Dashed and
    dotted lines show respectively the dependence of
    $\frac{\rho_{\phi}}{\rho_{\chi}}$ and $\frac{n_{\phi}}{n_{\chi}}$
    on $\alpha$. Extrapolation gives $\rho_{\phi} \simeq \rho_{\chi}$
    at $\alpha \simeq 3\times 10^{-8}$. LatticeEasy parameters are as
    in Fig.~(\ref{fig1}). For a physical value of the Higgs
    self-coupling $(\lambda\simeq 0.1$ the energies become comparable
    even closer to $\alpha=10^{-7}$.}
  \label{fig3}
\end{figure}

On Fig.~\ref{fig3} one can see the dependence of the total energy
transferred into the Higgs field as a function of the inflaton--Higgs
coupling $\alpha$. One can draw the conclusion that parametric
resonance effects only become important at $\alpha\sim 10^{-7}$, which
is too large a value.  Thus, the reheating process proceeds by means
of the simple decay of the inflaton (generated abundantly by
parametric resonance) into the Higgs particle.  This process will be
analysed \emph{analytically} in the next subsection, where we will
advocate that this perturbative inflaton decay really reheats the
Universe at lower values of the parameter $\alpha$.

%%%%%%%%%%%%%%%%%%%%%%%%%%%%%%%%%%%%%%%%%%%%%%%%%%%%%%%%%%%%%%%%%%%%%%%%
%%%%%%%%%%%%%%%%%%%%%%%%%%%%%%%%%%%%%%%%%%%%%%%%%%%%%%%%%%%%%%%%%%%%%%%%

%%%%%%%%%%%%%%%%%%%%%%%%%%%%%%%%%%%%%%%%%%%%%%%%%%%%%%%%%%%%%%%%%%%%%%%%
%%%%%%%%%%%%%%%%%%%%%%%%%%%%%%%%%%%%%%%%%%%%%%%%%%%%%%%%%%%%%%%%%%%%%%%%

\subsection{Light inflaton case ($m_I<2m_H$)}

While the parametric resonance regime for the Higgs is terminated
quite early, the fluctuations of the inflaton field continue to grow
exponentially.  Since the amount of the energy transferred into the Higgs
field is practically negligible the dynamics of the inflaton field is
very close to that of the pure quartic inflaton model which was
analyzed numerically in \cite{Micha2003,Micha2004}.  In brief, the
inflaton zero mode keeps driving the exponential grows of the nonzero
modes until roughly half of its energy is transferred into the
inflaton particles.  After that the re-scattering processes become
important, slowly moving the inflaton particle distribution to thermal
equilibrium.  At some moment the scattering process $2I\to 2H$ becomes
important and the Higgs particle (together with all other SM
particles) is generated and the standard thermal history of the
Universe takes over.  The easiest way to estimate the equilibration
temperature of this process is to compare the mean free path
$n\sigma_{2I\to 2H} \sim n\frac{\alpha^2}{\pi p_\mathrm{avg}^2}$,
where $p_\mathrm{avg}$ is the average inflaton momentum, with the
Hubble expansion rate\footnote{$\mPl=\unit[2.44\times10^{18}]{GeV}$ is
  the \emph{reduced} Planck mass.}  $H=\frac{T^2}{m_{\rm
    Pl}}\sqrt{\frac{\pi^2g_*}{90}}$.  For the thermal distribution of
the inflaton particles this leads to the estimate
\begin{equation}
  \label{eq:Trh}
  T_R \approx
  \frac{\zeta(3)\alpha^2}{\pi^4}{\sqrt{90 \over g_*}}m_{\rm Pl}
  \;,
\end{equation}
However, the distribution of the inflaton excitations may be,
generally, rather far from thermal equilibrium
\cite{Micha2003,Micha2004}.  Evolution of the occupation numbers of
the inflaton modes was found to be self similar in
\cite{Micha2003,Micha2004}
\begin{equation}
  \label{eq:1}
  n(k,\tau)=\tau^{-q}n_0(k\tau^{-p})
  ;,
\end{equation}
where $\tau$ is the conformal time, $k$ is the comoving momentum, and
$p=1/5$ for three particle interactions and $1/7$ for four particle
interactions, $q\sim 4p$.  The only relevant for us property of the
function $n_0(k\tau^{-p})$ is that the average momentum in
\eqref{eq:1} at the beginning of reheating after inflation is
$\beta^{1/4}\Mpl$.  Thus, the average momentum at later times is
smaller than expected from the total energy density,
$p_\mathrm{avg}/T\sim (\Mpl/T)^p\beta^{(1+p)/4}$, where
$T\sim\rho^{1/4}$ is now not a real temperature, but rather a
parameter defining the energy density\footnote{After thermalization
  into the SM particles $T$ transforms into the real temperature, up
  to the change of the number of d.o.f.} (cf.\ equilibration time
description in \cite{Micha2003,Micha2004}).  This enhances the
$2I\to2H$ cross section together with the $I$ number density,
increasing the estimate \eqref{eq:Trh} by a factor
$(T/p_\mathrm{avg})^3$.  This leads to the increase of the equilibration
temperature by a factor $10^5$ for four particle interaction, $p=1/7$,
and by a factor $10^2$ for three particle interaction, $p=1/5$.  Exact
calculation of the equilibration temperature requires extensive
numerical study, but, in any case, the expression \eqref{eq:Trh}
should be considered as the lower bound, while $10^5 T_R$ is the upper
(most conservative) bound.

Requiring that $T_{R}>\unit[150]{GeV}$ we can obtain the lower bound on
$\alpha$
\begin{equation}
  \label{alpha1}
  \alpha\ge 7.3\times 10^{-8}
  \;,
\end{equation}
for the thermal estimate \eqref{eq:Trh} and
\begin{equation}
  \label{eq:alpha1conserv}
  \alpha\ge 7\times 10^{-10}
  \;,
\end{equation}
for the most conservative estimate of non-thermal distribution of the
inflaton.\footnote{Strictly speaking, one should also check if there is any
kinematical suppression of the process. This may lead to ${\mathcal O}(1)$ corrections and is, in fact,
beyond the precision of present estimates.}

While the bound (\ref{alpha1}) roughly coincides with the one at which
the energy transfer to the Higgs field becomes effective enough to
significantly deplete the zero mode of the inflaton (see
Fig.~\ref{fig3}) while the value given by (\ref{eq:alpha1conserv}) is
about two orders of magnitude smaller.  We can, therefore, conclude
that the upper bound on the inflaton mass is given by
\begin{equation}
  m_I \le
  (0.14\div1.40)\left(m_{H}\over 150\,\mathrm{GeV}\right)
  \sqrt{\beta\over 1.3\times 10^{-13}}\,\mathrm{GeV}
  \;,
\end{equation}
where the range corresponds to the thermal or the most conservative
non-thermal estimates.

%%%%%%%%%%%%%%%%%%%%%%%%%%%%%%%%%%%%%%%%%%%%%%%%%%%%%%%%%%%%%%%%
\subsection{Heavy inflaton case ($m_I>2m_H$)}

In this case the inflaton mass allows for the direct decay of the
inflaton into two Higgs particles.  The corresponding decay rate is
given by
\begin{equation}
  \Gamma(I\to 2H)
  =\frac{1}{2}\sqrt{\alpha^3\over 2\pi^2\beta}\,m_{H}
  =\frac{\beta}{8\pi}\frac{m_H^4}{m_I^3}
  \;.
\end{equation}
Comparing this rate with the Hubble parameter and requiring again for
the reheating temperature $T_R>\unit[150]{GeV}$ we get
\begin{equation}
  m_I<440 \left(\frac{m_H}{150\,\mathrm{GeV}}\right)^{4/3}
          \left({\beta\over 1.3\times 10^{-13}}\right)^{1/3}
          \mathrm{GeV}
          \;.
\label{heavy}
\end{equation}
Of course, in the case $\alpha\lesssim\beta/8$ the generation of the cosmological perturbations is different from the case of pure quartic inflation. The 
Higgs field becomes relatively light and the parameter space of the model is modified.  In particular, isocurvature fluctuations which one would generically expect in the two-field model have to be somehow suppressed. This will put the restriction on the allowed values $(\alpha,\beta)$. The analysis of this parameter space is very involved. One can expect, for example, that the parameter 
$\beta$ can differ from its value in the case of 
pure quartic inflation. That is one of the reasons why the 
parametric dependence on $\beta$ is kept in (\ref{heavy}).\footnote{Note, however, that the dependence of the bound (\ref{heavy}) on $\beta$ is rather mild.} 

%%%%%%%%%%%%%%%%%%%%%%%%%%%%%%%%%%%%%%%%%%%%%%%%%%%%%%%%%%%%%%%%%%%%%%%%
\section{WMAP constraints and non-minimal coupling}

Finally let us discuss the constraints on the model from the WMAP data
\cite{Spergel:2006hy}.  As was already mentioned in the inflationary
regime the model is indistinguishable from the pure quartic potential
inflation.  One should then confront the fact that the amplitude of
the tensor perturbations is too large.  One possible resolution of
this problem is to assume that the inflaton $\chi$ has non-minimal
coupling to gravity \cite{Tsujikawa:2004my}.  We will repeat here the
estimates following closely
\cite{Kaiser:1994vs,Bezrukov:2008cq,Bezrukov:2007ep}. We will take the
following action as an example
\begin{multline}
  \label{nonmin}
  S=\int d^4x \sqrt{-g}\,
  \Bigg[-\left({m^2+\xi\chi^2\over 2}\right)R\\
  +{1\over
    2}(\partial\chi)^2+|\partial\Phi|^2-V(\chi,\Phi)\Bigg]
  \;,
\end{multline}
where $m\simeq\Mpl$.  Even if the coupling $\xi$ is zero at a tree
level one can expect that it will be generated \emph{via} radiative
corrections.  As it will be discussed below even for small values of
$\xi$ the coupling $\beta$ will deviate from the one, obtained from
the COBE normalization in the absence of the non-minimal coupling
$\beta|_{\xi=0}\sim 1.3\times 10^{-13}$.

%%%%%%%%%%%%%%%%%%%%%%%%%%%%%%%%%%%%%%%%%%%%%%%%%%%%%%%%%%%%%%%%%%%%%%%%
\begin{figure}
  \psfrag{z}[bl]{$\beta$}
  \psfrag{y}[bl]{    $\xi$}
  \psfrag{k}[bc]{\scriptsize{0.001}}
  \psfrag{0.100}[bl]{\scriptsize{0.1}}
  \psfrag{0.050}[bl]{\scriptsize{0.05}}
  \psfrag{0.020}[bl]{\scriptsize{0.02}}
  \psfrag{0.010}[bl]{\scriptsize{0.01}}
  \psfrag{0.005}[bl]{\scriptsize{0.005}}
  \psfrag{0.002}[bl]{\scriptsize{0.002}}
  \includegraphics[width=0.95\linewidth]{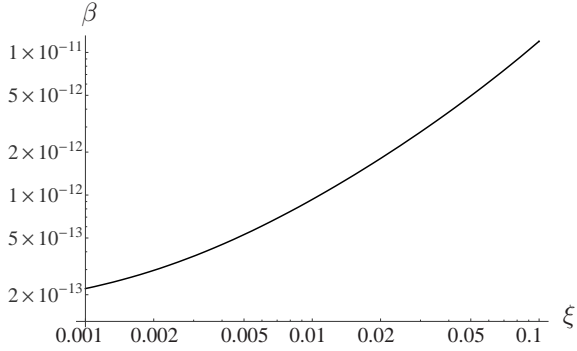}
  \caption{The dependence of the quartic coupling $\beta$ on the
    non-minimal coupling parameter $\xi$.}
  \label{fig4}
\end{figure}
%%%%%%%%%%%%%%%%%%%%%%%%%%%%%%%%%%%%%%%%%%%%%%%%%%%%%%%%%%%%%%%%%%%%%%%%

The bound on the tensor-to-scalar ratio comes from the perturbations
generated at $N\simeq 62$ e-foldings (see, e.g.\ \cite{Lyth:1998xn})
before the end of inflation.  In that regime the Higgs part of the
model is not important and can be dropped to simplify the
discussion. The inflaton part of (\ref{nonmin}) as it appears in
Jordan frame by means of the conformal transformation can be rewritten
as (hat denotes transformed quantities)
\begin{align}
  S_J=\int d^4 x\sqrt{-\hat g}\Bigg[&-{\Mpl^2\over 2}\hat R
  \notag\\
  &+{1\over
      2}(\partial\hat\chi)^2-U(\hat\chi)\Bigg]
  \;,
\end{align}
where
\begin{equation}
  \hat
  g_{\mu\nu}=\Omega^2g_{\mu\nu}
  \;,\quad
  \Omega^2 \simeq 1+{\xi\chi^2\over\Mpl^2}
  \;,
\end{equation}
and the new field $\hat\chi$ is defined as
\begin{equation}
  {d\hat\chi\over d\chi}=\sqrt{\Omega^2+6\xi\chi^2/\Mpl^2\over
    \Omega^4}
  \;.
  \label{nf}
\end{equation}
The new potential is given by
\begin{equation}
  U(\hat\chi)={\beta\over 4\Omega(\hat\chi)^4}\chi^4(\hat\chi)
  \;.
\end{equation}
We assume that $\xi\chi^2_\mathrm{e}/\Mpl^2\lesssim 1$, where $\chi_{\rm
  e}$ is the value of the inflaton field at the end of inflation, so
the contribution to the effective Plank mass vanishes after the
inflationary period.  In that case the dynamics of the model with the
action (\ref{nonmin}) after inflation is not different from that of
the $\nu$MSM model with the potential \eqref{pot}.  This suggestion
corresponds to $\xi<0.1$, see \eqref{condition}.  Following
\cite{Kaiser:1994vs,Tsujikawa:2004my} one can find that the first
slow-roll parameter $\epsilon$ is given by
\begin{equation}
  \epsilon={8\Mpl^4\over\chi^2(\Mpl^2+{\xi\chi^2(1+6\xi)})}
  \;.
\end{equation}
Slow-roll ends when $\epsilon=1$. From that one can find that
\begin{equation}
  \begin{split}
    {\xi\chi^2_\mathrm{e}\over \Mpl^2}=&{1\over
      2(1+6\xi)}\left(\sqrt{192\xi^2+32\xi+1}-1\right)\\&\approx
    8\xi+{\mathcal O}(\xi^2),~(\xi\ll 1)
    \;.
  \end{split}
  \label{condition}
\end{equation}

%%%%%%%%%%%%%%%%%%%%%%%%%%%%%%%%%%%%%%%%%%%%%%%%%%%%%%%%%%%%%%%%%%%%%%%%
\begin{figure}
  \begin{center}
    \includegraphics[width=\columnwidth]{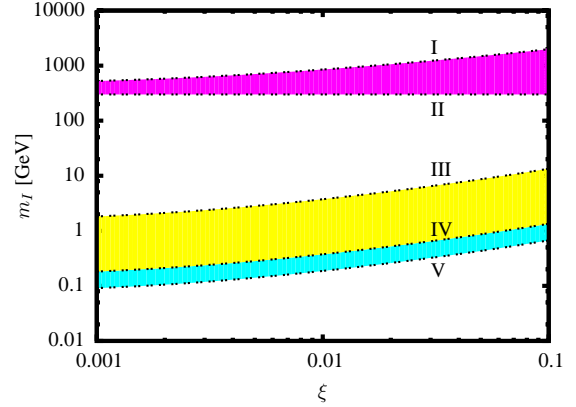}
  \end{center}
  \caption{Bounds on the inflation mass for successful reheating.
    Allowed regions for the case of $II\to HH$ scattering (lower
    region, light inflaton) and inflaton decay (upper region, heavy
    inflaton).  Higgs mass is chosen $m_H=\unit[150]{GeV}$ and
    $T_R\geq \unit[150]{GeV}$. Bounds are: I (inflaton decay), II
    ($m_I\geq2 m_H$), III (2-2 scattering, non-thermal $I$
    distribution), IV (2-2 scattering, thermal $I$ distribution), V
    ($\alpha\le 3\times 10^{-7}$, smallness of radiative
    corrections).}
  \label{fig6}
\end{figure}
%%%%%%%%%%%%%%%%%%%%%%%%%%%%%%%%%%%%%%%%%%%%%%%%%%%%%%%%%%%%%%%%%%%%%%%%

The number of e-foldings from the moment when the inflaton field has
the value $\chi_N$ till the end of inflation is given by
\begin{align}
  \label{ef}
  N & ={1\over\Mpl^2}\int^{\chi_N}_{\chi_\mathrm{e}}
  {U\over (dU/d\chi)}\left({d\hat\chi\over d\chi}\right)^2d\chi
  \\
  & ={1\over8}\left[\frac{\chi_N^2-\chi^2_\mathrm{e}}{\Mpl^2}(1+6\xi)
    -6\ln\left({\Mpl^2+\xi\chi_N^2} \over
      {\Mpl^2+\xi\chi^2_\mathrm{e}}\right)\right]
  \;.\notag
\end{align}
Since $\xi\ll 1$ one can find that with a good accuracy $\chi_N\approx
2\sqrt{2(N+1)\over 1+6\xi}m_{\rm Pl}$.  The tensor-to-scalar ratio is
given by \cite{Spergel:2006hy}
\begin{align}
  r\equiv 16\epsilon&={128\Mpl^4\over\chi^2_N(\Mpl^2+{\xi\chi^2_N(1+6\xi)})}
  \notag\\
  &\approx
    \frac{16(1+6\xi)}{(N+1)(8\xi(N+1)+1)}
  \;.
  \label{beta}
\end{align}
 
One can see \cite{Tsujikawa:2004my} that roughly in the interval
$\xi=0.001 \div 0.1$ this ratio satisfies the WMAP constraints.  The
value of the inflaton self-coupling as a function of $\xi$ can be
found from the COBE normalization
$U(\chi_N)/\epsilon(\chi_N)=(0.027\Mpl)^4$.  The corresponding
behavior is shown in Fig.~\ref{fig4}.  This introduces slight growth
of $\beta$ with $\xi$, and thus increases all bounds simultaneously,
which is demonstrated in Fig.~\ref{fig6}.

%%%%%%%%%%%%%%%%%%%%%%%%%%%%%%%%%%%%%%%%%%%%%%%%%%%%%%%%%%%%%%%%%%%%%%%%
\section{Conclusions}

In Fig.~\ref{fig6} we combined the bounds on the inflaton mass we have
found so far.  We can conclude, therefore, that the mass of the
inflaton in the $\nu$MSM inflation \cite{Shaposhnikov:2006xi} should
be roughly in the range
\begin{equation}
  \unit[0.1]{GeV}\lesssim m_I\lesssim \unit[10]{GeV}
\end{equation}
in the case when it is light and in the range
\begin{equation}
  \unit[300]{GeV}\lesssim m_I\lesssim \unit[1000]{GeV}
\end{equation}
in the case when the inflaton-Higgs coupling is very small.

These bounds could be evaded in models with arbitrary scalar field
potentials, but the fact of the strong lower bound from reheating on
the coupling between the inflaton and the Higgs should remain rather
universal.

Values of $\xi$ larger then $0.1$ (and, therefore larger lower and
upper bounds on the inflaton mass) are also allowed as well.  However,
since the dynamics of the model at preheating may be strongly modified
from the one we have studied in this Letter it is hard for us to make
any statements in that case, and we leave this for future analysis.

\section*{Acknowledgments}

We thank M. Shaposhnikov and D. Gorbunov for many helpful discussions.

%%%%%%%%%%%%%%%%%%%%%%%%%%%%%%%%%%%%%%%%%%%%%%%%%%%%%%%%%%%%%%%%%%%%%%%%
%\bibliography{all}
%\bibliographystyle{h-elsevier3}

\end{document}